\documentclass[12pt]{iopart}
\usepackage{placeins}

\usepackage{tikz}
\usepackage{graphicx}
\usepackage{epstopdf}
\usepackage{citesort}
\begin{document}

    \newcommand{\threepartdef}[6]
    {
        \left\{
        \begin{array}{lll}
            #1 & \quad \mbox{for } #2 \\
            #3 & \quad \mbox{for } #4 \\
            #5 & \quad \mbox{for } #6
        \end{array}
        \right.
    }

\title{Transient chaos in time-delayed systems subjected to parameter drift}

\author{Julia Cantis\'{a}n $^1$, Jes\'{u}s M. Seoane $^1$, Miguel A.F. Sanju\'{a}n $^1$ $^2$}
\address{$^1$ Nonlinear Dynamics, Chaos and Complex Systems Group, Departamento de F\'{i}sica, Universidad Rey Juan Carlos \\ Tulip\'{a}n s/n, 28933 M\'{o}stoles, Madrid, Spain}
\address{$^2$ Department of Applied Informatics, Kaunas University of Technology \\ Studentu 50-415, Kaunas LT-51368, Lithuania}

\begin{abstract}
External and internal factors may cause a system's parameter to vary
with time before it stabilizes. This drift induces a regime shift
when the parameter crosses a bifurcation. Here, we study the case of
an infinite dimensional system: a time-delayed oscillator whose time
delay  varies at a small but non-negligible rate. Our
research shows that due to this parameter drift, trajectories from a
chaotic attractor tip to other states with a certain probability.
This causes the appearance of the phenomenon of transient chaos. By using an ensemble approach, we find a gamma
distribution of transient lifetimes, unlike in other non-delayed systems where normal distributions have been found to govern the process. Furthermore, we analyze how the parameter change rate influences the tipping
probability, and we derive a scaling law relating the parameter value for which the
tipping takes place and the lifetime of the transient chaos with the
parameter change rate.
\end{abstract}

\maketitle

\section{Introduction}

A vast variety of systems are not static over time, in the sense that they can not be modeled by the same set of equations and parameters as time passes. External factors as the increase in
greenhouse gases in the context of climate dynamics
\cite{Kaszas2019a} are one of the possible causes. Also, this change
can be caused by internal factors, i.e., by the nature of the system
itself as in the case of some engineering systems that wear-out.
Furthermore, some systems may be modified manually by a
experimentalist that, for instance, injects chemical substances in a
reactor at a slowly varying flux \cite{Berglund1999}.
Mathematically, these processes are reflected by a drift in one of
the parameters of the model.

For systems suffering a drift at slow rates, the analysis of the
frozen-in system, that is, the model without time dependence in the
parameters, is of great importance. A vital part of this analysis
consists on the computation of bifurcation diagrams that depict the
dynamics of the frozen-in system for different values of a
parameter. It has been shown in \cite{Neishtadt1987,Neishtadt1988}
that there is a delay in the regime shift when a parameter drifts,
at small but non-negligible rates, and crosses a bifurcation point
($p_{bif}$). This transition occurs at a value of the parameter $
p_{cr}>p_{bif} $. This is usually referred to as the delay effect
in systems with parameter drift. It was also shown that the value of
$ p_{cr} $ depends on the parameter change rate. This type of
bifurcations that are crossed due to the time dependence of one of
the parameters are called dynamic bifurcations \cite{Benoit1991}.

Previous work on dynamic bifurcations has been initially restrained to
systems where only regular attractors were involved. More recently, some attention has been focused on chaotic attractors by using either maps \cite{Maslennikov2013,Maslennikov2018} or flows \cite{Cantisan2020a}. In both cases,
an ensemble approach is needed and a normal distribution was found
for the values of $ p_{cr} $. Here, we aim to broaden systems currently
explored, by studying dynamic bifurcations in an infinite
dimensional system: a time delayed oscillator. As far as we know,
this type of systems have not received much attention in the context
of dynamic bifurcations and they are worth a study due to the
ubiquity in Nature of time-delayed systems. Physically, the time delay
accounts for the finite propagation time of information. In the case
of laser arrays, for instance, this is due to the finite speed of
light \cite{Soriano2013} and the time delay determines its stability
\cite{Topfer2020}. Here, we consider that the time delay slowly
drifts with time and we explore its implications. This drift may
account, for instance, for the seasonal fluctuations to the
regeneration time of a resource \cite{Bartuccelli1997}.

Besides the delay effect, the parameter drift provokes another
interesting phenomenon: bifurcation-induced tipping
\cite{Ashwin2012}. When the system presents multistability
trajectories, for $ p>p_{cr} $, have to tip to any of the attractors
that are stable with a certain probability. This tipping probability
also depends on the parameter change rate. Finally, in a time
framework, we may consider that after the tipping, the drifting
system is in its steady state. Before that, it is in its transient
state. Depending on the parameter change rate, the transient state
would last for a longer or shorter period of time. This may be a problem in some engineering systems as a parameter drift would be identified long after its start, possibly causing parameter drift failure \cite{Stevens2002,Wu2015}.

This article is organized as follows. In section \ref{sec1} we
present our system: the Duffing oscillator with time delay, and we
analyze its dynamics when the parameters are frozen-in. In section
\ref{sec2} we introduce the time dependence in the time delay
parameter. We explore the tipping probabilities in section
\ref{sec3} and the delay effect in section \ref{sec4}, deriving a
scaling law that relates the delay and the parameter change rate. Finally, the time framework is analyzed in section \ref{sec5}, where the
arising transient chaos phenomenon is characterized. Discussions
and conclusions are drawn at the end.

\section{Time-delayed Duffing oscillator} \label{sec1}

As a paradigmatic example of a nonlinear oscillator with delay, we
study the transient dynamics of the undamped and unforced Duffing oscillator with
a delay term of the form $ \gamma  x(t - \tau) $, where $ \tau $ is
the time delay and $ \gamma $ is the amplitude of the delay. The
equation for the system reads:
\begin{equation}
    \ddot{x} + \alpha x + \beta x^{3} + \gamma  x(t - \tau)=0.
    \label{system_eq}
\end{equation}

We consider the system in absence of dissipation since we focus our
study in the effect of the time delay term. This can
be performed experimentally if we consider that the experiments are carried out in the laboratory under very low pressure and
therefore in the situation of vacuum. In this physical context,
thanks to the time delay and the absence of damping, the
oscillator presents a region of parameters for which a chaotic
attractor is stable. We choose the same parameter values considered
in \cite{Rajasekar2016a}, that is, $ \alpha=-1 $, $ \beta=0.1 $ and
$ \gamma=-0.3 $. With these values the potential has the shape of a
double well, with an unstable fixed point at the origin and two
stable fixed points at the bottom of the wells. Taking $ x(t)=x(t-
\tau)=x^{*} $, we find the numerical values for the fixed points as
\begin{equation} \label{eq_x}
    x^{*}=0 \\ x^{*}_{\pm}=\pm\sqrt{\frac{-\alpha-\gamma}{\beta}}=\pm\sqrt{13}=\pm3.606.
\end{equation}

In this section, we present the system dynamics for the frozen-in
case, that is, for the system when the time delay does not vary with
time. For different but fixed in time values of $ \tau $, the
system's behavior changes. To uncover this, we explore the
attractors that appear for two values of $ \tau $ that correspond to
the starting and ending scenarios of the ramp $ \tau(t) $ that is
explored in the next section.

\begin{figure}[h]
    \begin{center}
        \includegraphics[width=0.45\textwidth ]{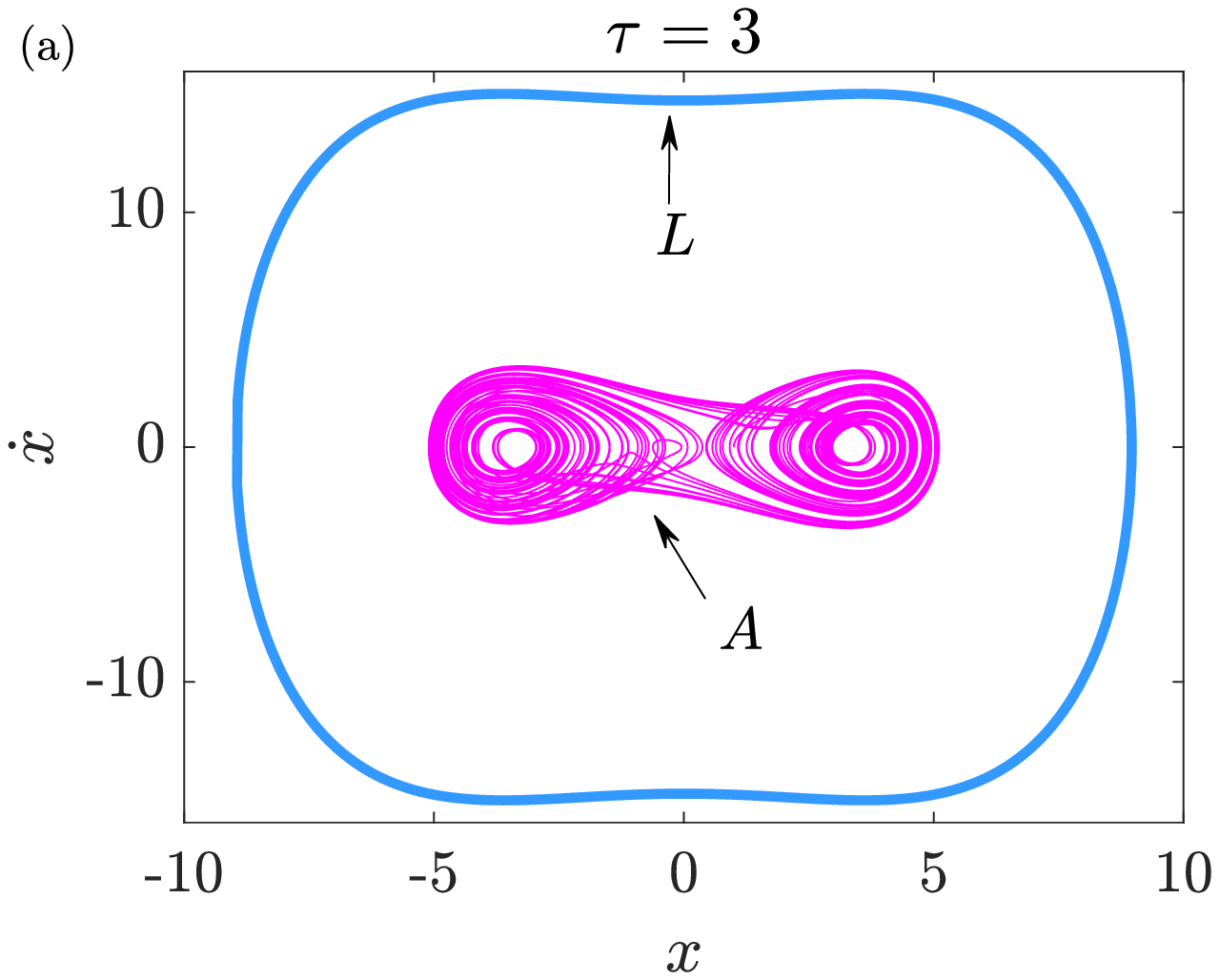}
        \includegraphics[width=0.45\textwidth]{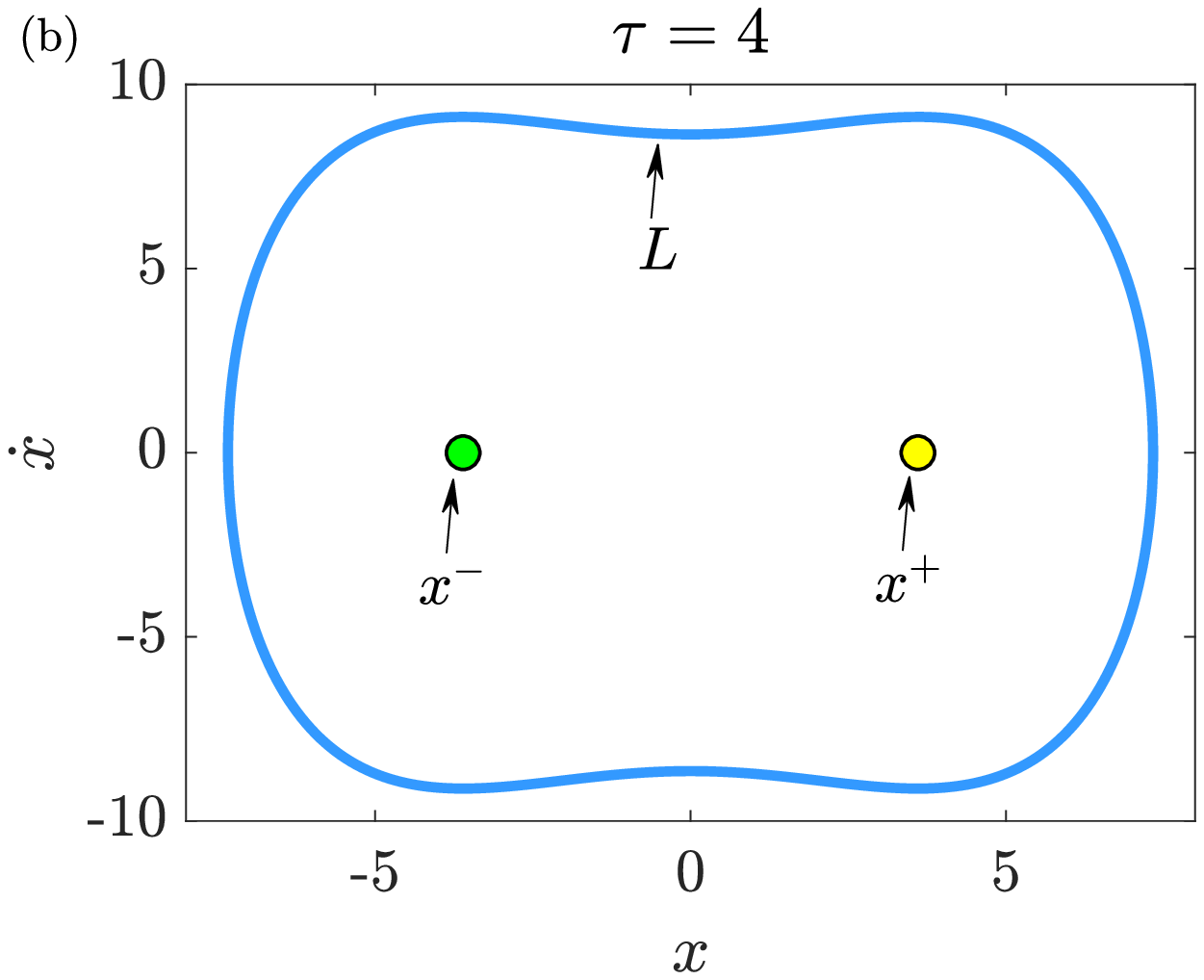}
    \end{center}
    \caption{ Attractors before and after the bifurcation at $ \tau=3.6 $. (a) For $ \tau=3 $, two attractors coexist: a limit cycle that we call $ L $ and a chaotic attractor, $ A $. (b) For $ \tau=4 $, the limit cycle is stable although its amplitude decreases with $ \tau $, as can be observed by noting the different scales of both figures. Furthermore, two new attractors appear: two symmetrical fixed points, $ x^{\pm} $.}
    \label{attractors}
\end{figure}

In the context of time-delayed systems, initial conditions are replaced by history functions: $ u_{0}(t)=(x_{0}, \dot{x_{0}}) $, which are a set of initial conditions in the continuous time interval $ [-\tau,0] $. The history functions that we use here are, for simplicity, constant functions. Due to the multistability of the system, different history functions lead to different attractors. In figure \ref{attractors}, we show all the coexisting attractors in phase space for $ \tau=3 $ and $ \tau=4 $.

In figure \ref{attractors} (a), we observe two attractors: a limit
cycle, $ L $, and a chaotic attractor, $ A$. The bifurcation diagram
for $ \tau \in (0, 5] $ and history functions $ u_{0}=1 $ and $
u_{0}=-1 $, which was calculated in  \cite{Cantisan2020}, showed
that the chaotic attractor disappeared at $ \tau=3.6 $.
Consistently, for $ \tau=4 $, in figure \ref{attractors} (b), it can
be seen that the chaotic attractor is no longer stable and that
there are two new attractors, corresponding to the previously
calculated fixed points, $ x^{+} $ and $ x^{-} $ from equation (\ref{eq_x}).
Although the limit cycle is still stable in both scenarios, it is
important to remark that its amplitude decreases with $ \tau $. This
fact can be noticed by checking the different scales on the axis in
figures \ref{attractors} (a) and (b).

Now, we calculate the basins of attraction for $ \tau=3 $ and $
\tau=4 $. For this purpose, we distribute $ N=103041 $ constant
history functions uniformly in a grid of $ [-8, 8] \times [-8, 8] $, in steps of $0.05$. The basins are computed integrating (\ref{system_eq}) for every
history function with the algorithm Tsit5 from Julia
\cite{Bezanson2017} for delay differential equations, which is a
Runge-Kutta method with adaptative stepsize-control. The parameters
for the numerical calculation have to be taken cautiously as the
results have to be very precise to lead to the correct attractor,
specially in the chaotic region. In our case, we took RelTol=$
10^{-9} $, AbsTol=$ 10^{-12} $ and maxiters=$ 10^{9} $. The
criterion for convergence to fixed points is that the trajectory
enters a square of $ 0.01 \times 0.01 $. In the case of the limit
cycle, the criterion for convergence is that the peaks from the time
series become equally spaced, thus periodic. When none of these
criteria are met, the chaotic attractor is assigned.

\begin{figure}[h]
    \begin{center}
        \includegraphics[width=0.45\textwidth]{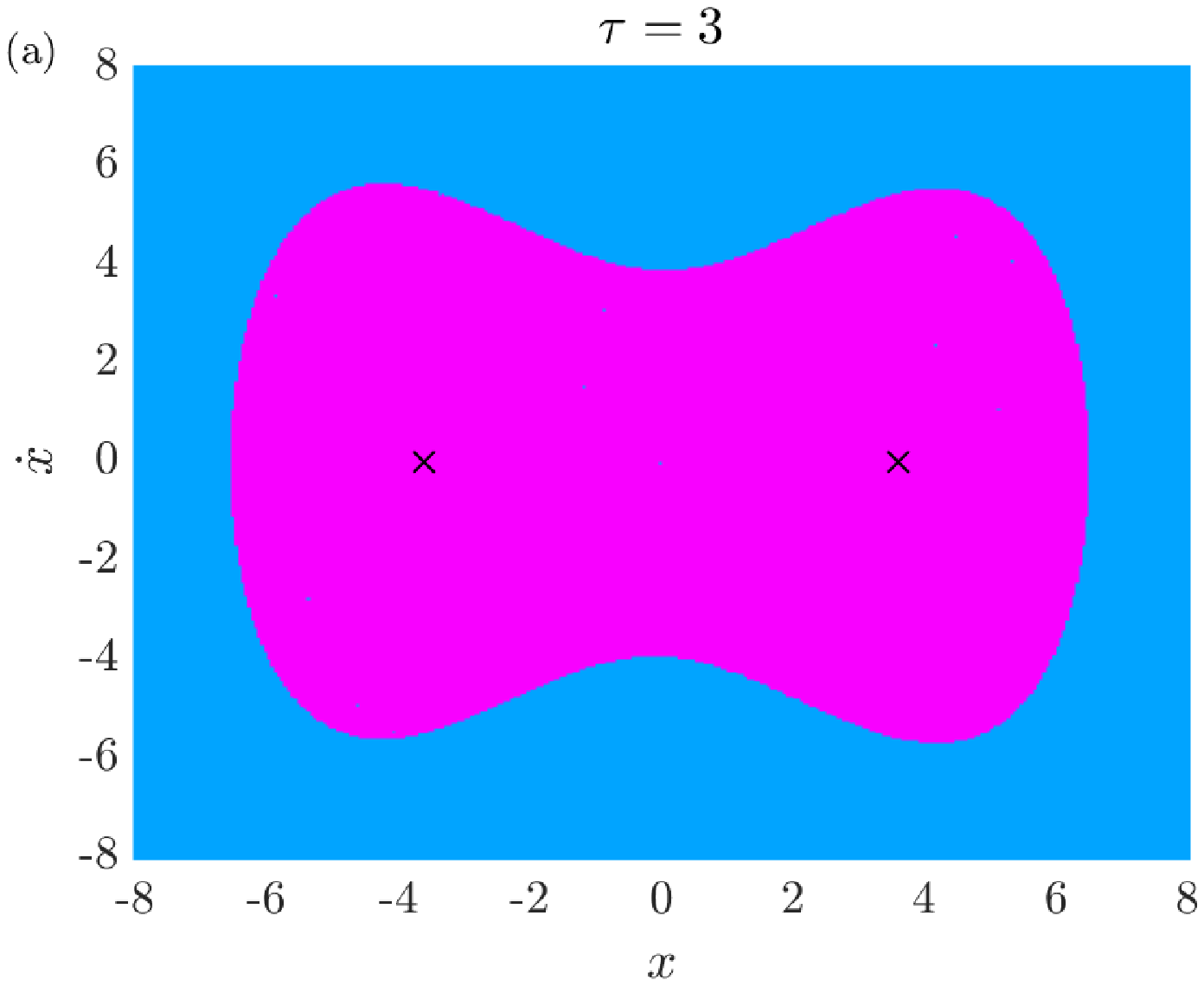}
        \includegraphics[width=0.45\textwidth]{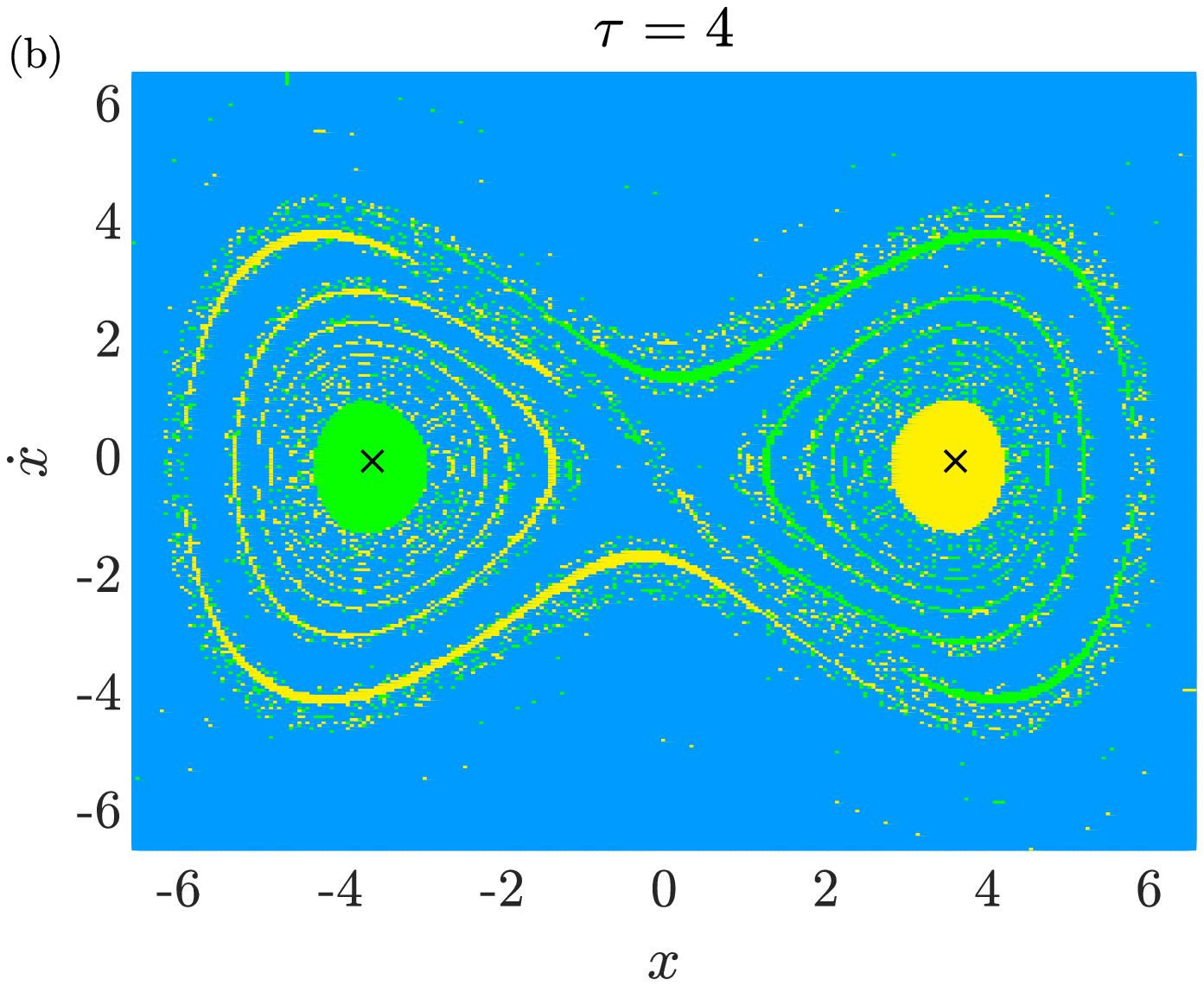}
    \end{center}
    \caption{Basins of attraction before and after the bifurcation at $ \tau=3.6 $. Two black crosses point the $ x^{\pm} $ attractors. (a) For $ \tau=3 $, the pink and bone-shaped region corresponds to the set of history functions that go the the chaotic attractor. On blue, the ones that go to the limit cycle. (b) For $ \tau=4 $, the bone shaped structure is replaced by the almost circular basins of $ x^{+} $ in yellow and $ x^{-} $ in green around the attractors, which are also intermingled with the blue basin maintaining the bone-shaped structure.}
    \label{basin_tau3y4}
\end{figure}

The basin of attraction for $ \tau=3 $ is depicted in
\ref{basin_tau3y4} (a). In blue, the history functions that lead to
the limit cycle, $ L $, and in pink the history functions that lead
to the chaotic attractor, $ A $. The fixed points $ x^{+}/x^{-} $
are also represented as black crosses. As it can be
seen, there is a bone-shaped region that corresponds to the chaotic
attractor basin. Outside this bone-shaped structure, all the
trajectories approach the limit cycle.

Following a similar sketch, the basin of attraction for $ \tau=4 $ is depicted in \ref{basin_tau3y4} (b). The yellow and green regions correspond to the set of history functions that lead to $ x^{+}$ and $x^{-} $ respectively. As it can be seen, around each fixed point there is a circular region of points with smooth boundaries that leads to them, in other words, trajectories that start near $ x^{\pm}= \pm (\sqrt{13},0) $ end up in the yellow/green attractor. Also, there are other history functions, mainly inside the previous bone-shaped structure, that approach $ x^{\pm} $.

\section{Time dependent delay, $ \tau(t) $} \label{sec2}

Once the behavior of the system for fixed values of the time delay
(before and after the bifurcation at $ \tau=3.6 $) is known, we let
this parameter slowly evolve with time between those scenarios. The
main difference with the previous section is that now we are
interested in the case of the parameter variation during the evolution
of the system. In other words, the parameter turns into a
slowly varying function of time of the form $
\tau(t)=\tau_{0}+\varepsilon t $, where $ \varepsilon $ is a
sufficiently small parameter compared to the natural time scale of
the system.

Dynamic bifurcations, as mentioned before, are the bifurcations
 which are crossed due to the time dependence of this parameter \cite{Benoit1991}. Here, we aim to study this
phenomenon for systems where not only chaotic attractors are
involved, but time delay is present too. This is an interesting area
as models that include a time delay cover the unavoidable phenomenon
of the finite propagation of information. We let not just a normal
multiplicative or additive parameter vary with time, but the time
delay itself.

The first difference in this analysis from the one of regular
attractors is that single trajectories are no longer representative
and do not contain all the possible dynamics. Thus, we follow an
ensemble of trajectories starting on the same set of history
functions used in the previous section to calculate the basins of
attraction.

Furthermore, we ought to redefine our system including the time delay dependence with time. We replace $ \tau $ in equation (\ref{system_eq}) by:
\begin{equation}
    \tau = \threepartdef { \tau_{0} } {t<t_{1}} {\tau_{0}+\varepsilon \cdot (t-t_{1})} { t_{1}<t<t_{2}} {\tau_{0}+\varepsilon \cdot (t_{2}-t_{1})} {t>t_{2}},
    \quad \quad
    \begin{tikzpicture}[baseline]
        \draw[ultra thick] (0,-1) -- (-1.5,-1)  node[anchor=east] {$ \tau_{0} $};
        \draw[ultra thick] (0,-1) -- (1.5,1);
        \draw[ultra thick] (1.5,1) -- (3,1);
        \draw[ dotted] (0,1.5) -- (0,-1.5)  node[anchor=north] {$ t_{1} $};
        \draw[ dotted] (1.5,1.5) -- (1.5,-1.5) node[anchor=north] {$ t_{2} $};
    \end{tikzpicture}
    \label{tau(t)}
\end{equation}
where $ \tau_{0} $ is the initial value of $ \tau $, $ t_{1} $ is the time for which the parameter shift starts and $ t_{2} $ when it ends. We refer to this system as the non-autonomous system, in contrast with the frozen-in system.

For simulation purposes, we take $ \tau_{0}=3 $ and we let $ t_{1}=100 $. This way, we let the system evolve to its steady state (remember figure \ref{basin_tau3y4} (a)), before the shift in $ \tau $ starts. The orbits evolve as in the previous section until $ t=t_{1} $, when $ \tau $ starts shifting and finally they reach one of the attractors that are stable after the bifurcation (see figure \ref{basin_tau3y4} (b)). For $ t_{2} $, we choose sufficient and different values depending on $ \varepsilon $ so that all the trajectories have the time to reach one of the attractors.

\section{Random tipping} \label{sec3}

\begin{figure}[h]
    \begin{center}
        \includegraphics[width=0.6\textwidth]{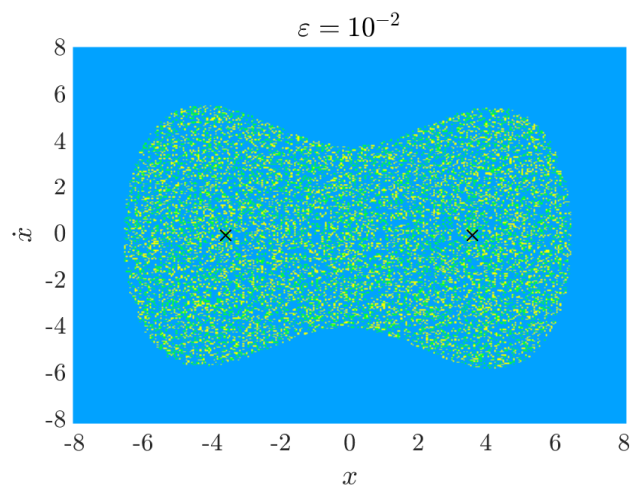}
    \end{center}
    \caption{State dependent basin of attraction for $ \varepsilon=10^{-2} $. As the chaotic attractor looses stability at $ \tau=3.6 $, the trajectories tip to one the attractors that are stable ($ L, x^{\pm} $). This leads to phenomenon of random tipping, and as result we obtain a riddle basin inside the bone-shaped structure.}
    \label{basin_tau3_eps102}
\end{figure}

Using the same terminology as in \cite{Kaszas2019}, we refer to the basin for the non-autonomous system as  scenario-dependent basin, because it depends on the parameters of the parameter shift. Figure \ref{basin_tau3_eps102} depicts the scenario-dependent basin of attraction for $ \varepsilon=10^{-2} $. In this case, $ t_{2}=350 $.

We distinguish three different basins corresponding to the attractors past the bifurcation, see figure \ref{attractors} (b). Outside the bone-shaped structure, all initial functions approach the limit cycle. If we compare this with figure \ref{basin_tau3y4}, we observe that the trajectories initially in the limit cycle never jump to a different attractor, not even for faster parameter change rates. Inside the bone-shaped structure, the yellow/green smooth boundary regions present for the frozen-in system disappear now, even if we stop the shift at  $ \tau=4 $. Due to the presence of the chaotic attractor, there is no pattern, and the history functions that lead to each basin are completely intermingled. This is because the final destination of each trajectory depends on the precise moment that it is caught wandering in the chaotic attractor when it tips. We may say that the chaotic attractor acts as a memory-loss agent. In other words, predictability of individual trajectories is lost because the passage through the chaotic attractor induces fractal basins of attraction. Similar phenomena \cite{Kaszas2019} for non time-delayed systems has been addressed as a random tipping.

We may repeat the same procedure for different values of the
parameter rate of change, $ \varepsilon $. In table \ref{table} we
show how the number of trajectories that approach each attractor
varies with $ \varepsilon $. This is also referred to as tipping
probability as it reflects the probability of a random history
function to end in each of the attractors.

\begin{table}
    \caption{\label{table}Percentage of trajectories that approach each attractor depending on the value of the parameter change rate.}
    \begin{indented}
        \item[]\begin{tabular}{@{}llll}
            \br
            $ \varepsilon $& $ L \quad (\%)$& $ x^{+} \quad (\%) $& $ x^{-} \quad (\%)$\\
            \mr
            $ 10^{-2} $&$ 70.30   $&$ 14.85  $&$ 14.85  $\\
            $ 7 \cdot 10^{-3} $& $ 74.48 $ & $ 12.76 $ & $ 12.76 $ \\
            $ 5 \cdot 10^{-3} $& $ 80.02 $ & $ 9.99 $ & $ 9.99 $ \\
            $ 3 \cdot 10^{-3} $& $ 88.58 $ & $ 5.71 $ & $ 5.71 $ \\
            $  10^{-3} $& $ 99.05 $ & $ 0.475 $ & $ 0.475 $ \\
            \br
        \end{tabular}
    \end{indented}
\end{table}

\begin{figure}[h]
    \begin{center}
        \includegraphics[width=0.6\textwidth]{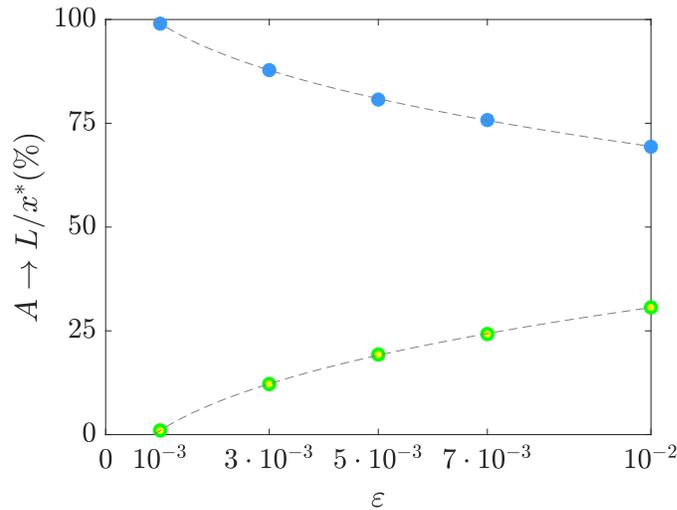}
    \end{center}
    \caption{Dependence on the parameter change rate for the percentage of trajectories that leave the chaotic attractor and approach the limit cycle (blue points) or the fixed points, $ x^{\pm} $ (yellow/green points). For higher parameter change rates, more trajectories end up in the limit cycle.}
    \label{percentage}
\end{figure}

Due to the symmetry of the potential well, the percentages for
attractors $ x^{+} $ and $ x^{-} $ are the same. For faster changes
in $ \tau $, the probability that a trajectory ends up in the fixed
points $ x^{\pm} $ rather than in the limit cycle, $ L $ grows. For
the sake of clarity, this is also represented in figure
\ref{percentage}, where the blue points correspond to the percentage
of trajectories that leave the chaotic attractor, $ A $, and end up
in the limit cycle, $ L $. In the same way, the yellow/green points
correspond to the percentage of trajectories that leave the chaotic
attractor, $ A $, and end up in $ x^{\pm} $.

\section{Scaling law} \label{sec4}

Although we have uncovered the scenario-dependent basins of attraction, one question is still lacking. When does the orbit jump from the chaotic attractor to either of the attractors that are stable at the end of the parameter shift? From \cite{Cantisan2020}, we know that the bifurcation through which the chaotic attractor looses stability occurs at $ \tau=3.6 $ for the frozen-in system. However, the picture here is different as our parameter varies during the evolution of the system. Now, we explore if the value of $ \tau $ for this transition, which we call $ \tau_{cr} $, is different from the one in the frozen-in case.

\begin{figure}[h]
    \begin{center}
        \includegraphics[width=0.45\textwidth ]{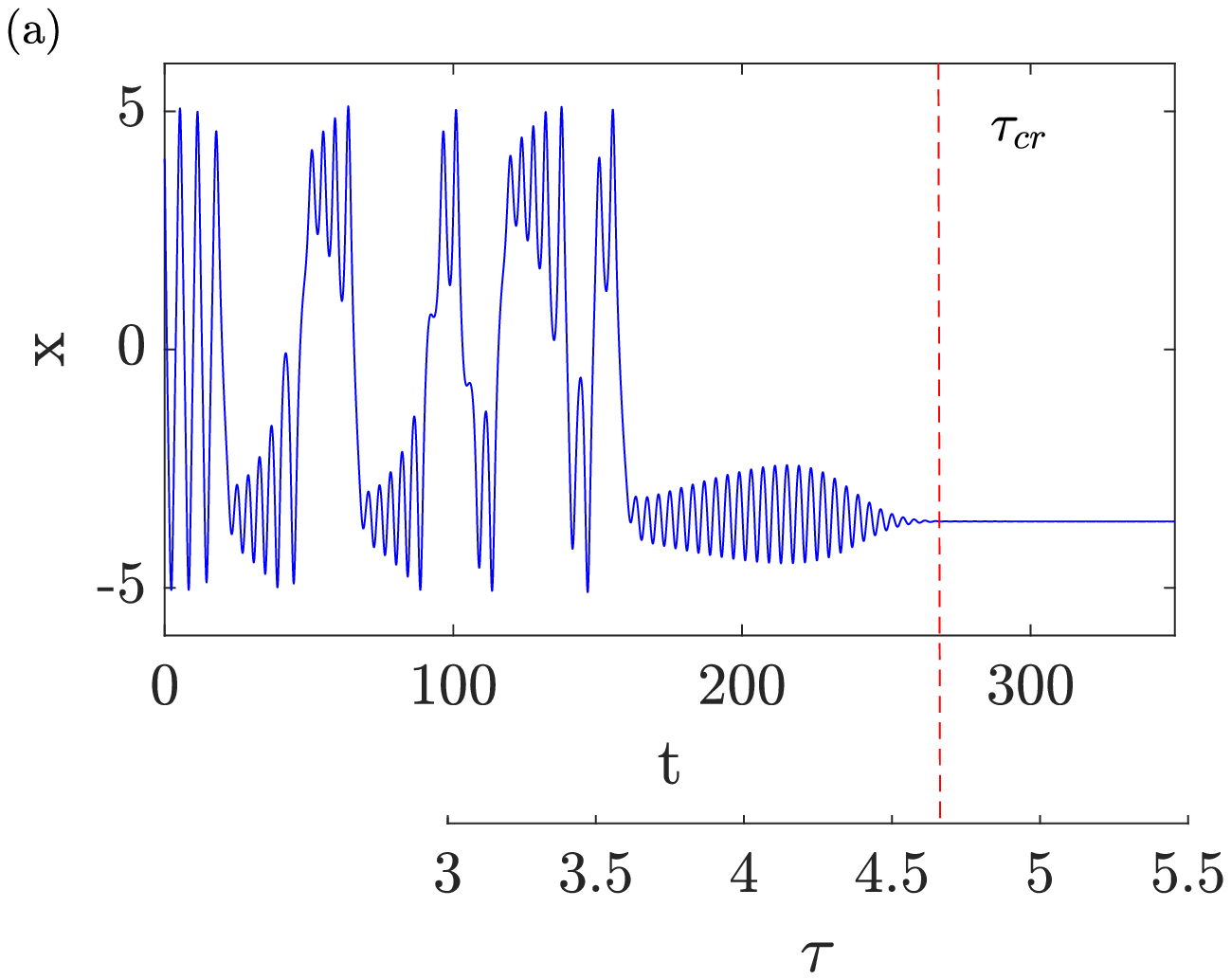}
        \includegraphics[width=0.45\textwidth]{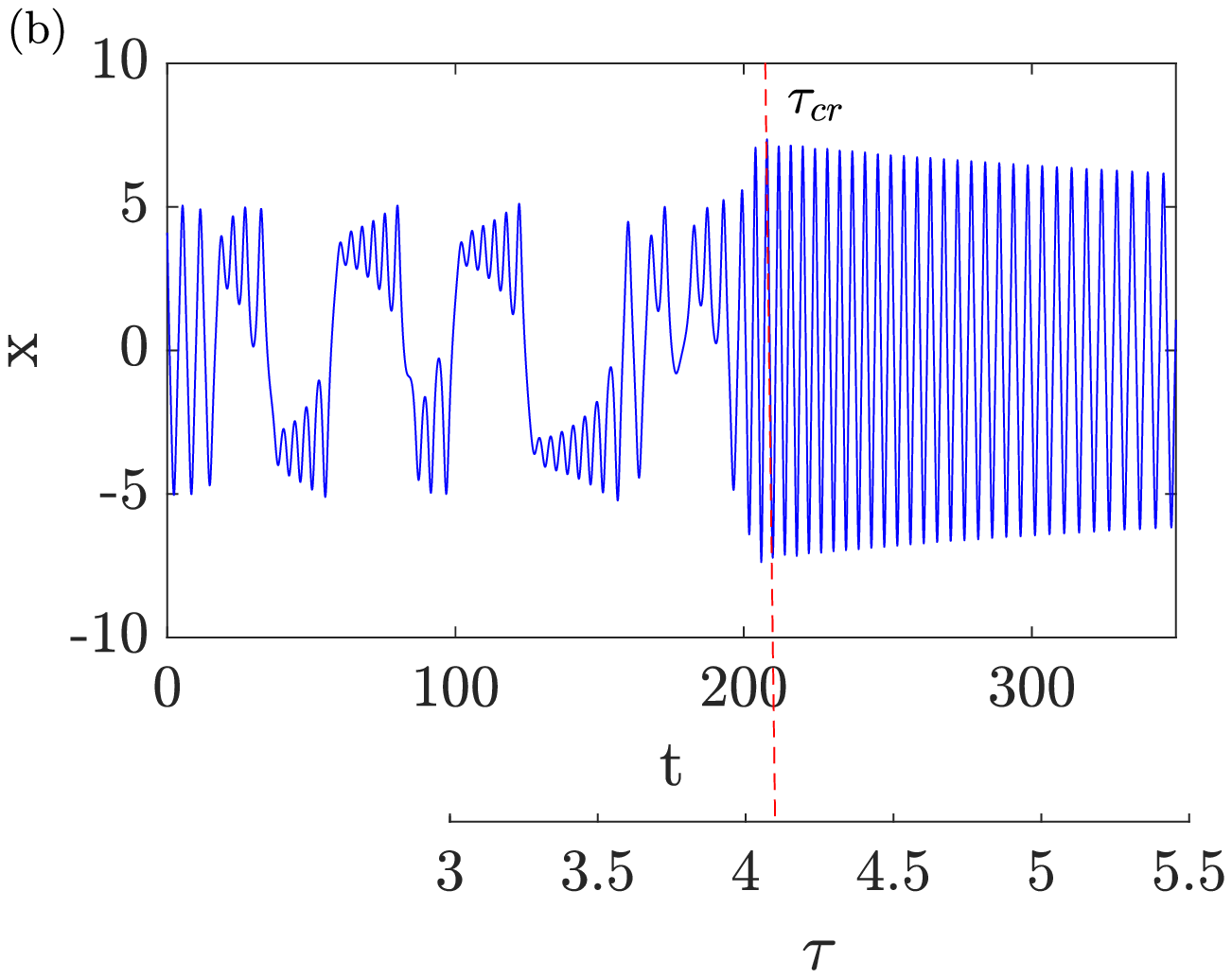}
    \end{center}
    \caption{Time series for the non-autonomous Duffing oscillator  with $ \varepsilon=  10^{-2} $. The secondary $x$-axis shows the time dependence in the $ \tau $ parameter. It can be seen that the transient lasts for a long period of time and that the transition to the steady state starts at a value of $ \tau $ past the bifurcation value for the frozen-in system, this is, $ \tau=3.6 $. This value ($ \tau_{cr} $) is marked with a vertical dotted line.}
    \label{tau_cr}
\end{figure}

In figure \ref{tau_cr}, we show two time series corresponding to two different history functions for the non-autonomous system with $ \varepsilon= 10^{-2} $. The secondary x-axis marks the evolution of the parameter. As it can be seen, in both cases, the behavior at first is chaotic. At $ t=100 $, $ \tau $ starts increasing. At $ \tau=3.6 $ (or $ t=160 $), the chaotic attractor lost stability for the frozen-in case. However, the chaotic behavior is still present for a while after until the system jumps to $ x^{-} $ in (a) or $ L $ in (b). The vertical dotted lines indicate this moment and the value of $ \tau_{cr} $. Between $ \tau=3.6 $ and $ \tau_{cr} $, the chaotic attractor is a metastable state of the system.

For numerical purposes, we calculated $ \tau_{cr} $ as the value of $ \tau $ for which the orbit is at a distance of $ dx=0.001 $ from $ x^{\pm} $, or the value of $ \tau $ for which the period of oscillation stabilizes and the amplitude starts decreasing.

\begin{figure}[h]
    \begin{center}
        \includegraphics[width=0.45\textwidth]{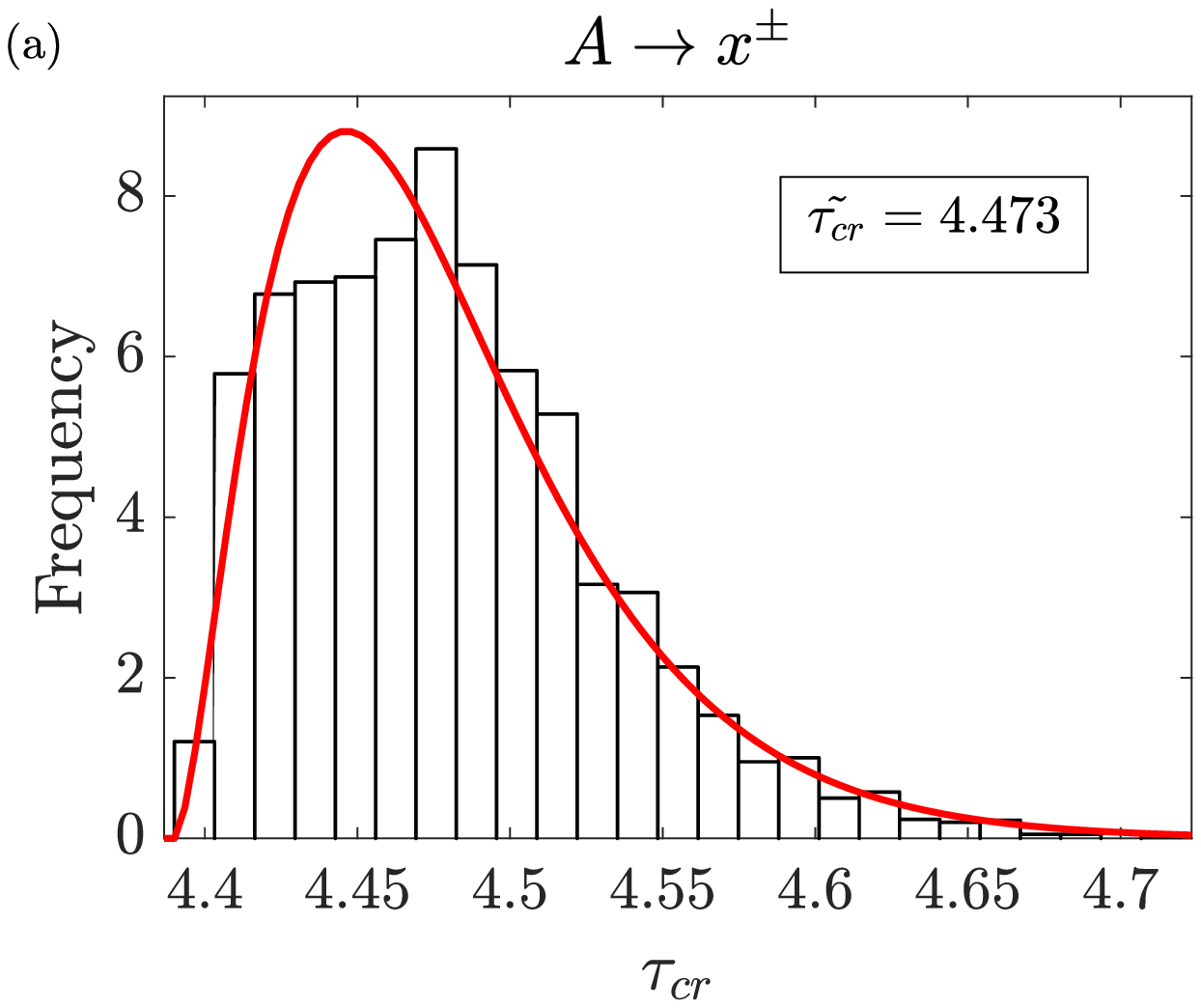}
        \includegraphics[width=0.45\textwidth]{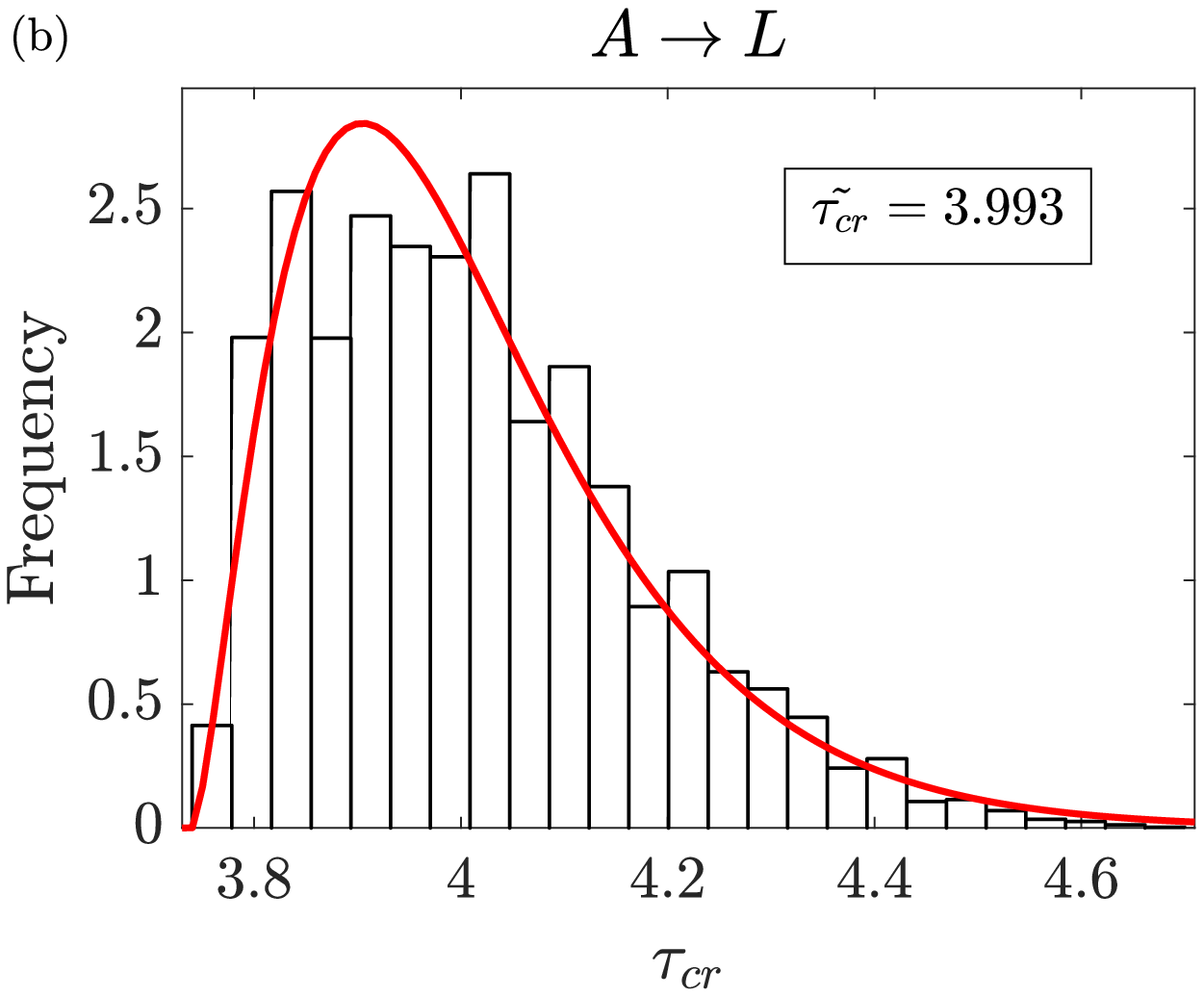}
    \end{center}
    \caption{Gamma distribution of $ \tau_{cr} $ for $ \varepsilon=7 \cdot 10^{-3} $ for the trajectories that leave the chaotic attractor and approach (a) the fixed points or (b) the limit cycle. The variance is higher in the later case, but the median value is smaller. Thus, history functions that end up in the limit cycle, on average leave the chaotic attractor before and for a wider range of values of $ \tau $.}
    \label{histogram}
\end{figure}

As previously mentioned, due to the chaotic attractor, this analysis has to be performed for a large set of history functions. This is why for each value of the parameter change rate, we do not obtain only one value of $ \tau_{cr} $; instead we obtain a distribution of values.  In figure \ref{histogram} we can see the form of this distribution for the cases $ A \rightarrow x^{\pm} $ and $ A \rightarrow L $ when $ \varepsilon=7 \cdot 10^{-3} $. Both of them are fitted to a gamma distribution with positive skew. It is interesting to notice that the variance is more pronounced for the trajectories going to the limit cycle, while for the ones going to $ x^{\pm} $, the deviation in the values for the transition is smaller. Inside both histograms, we have indicated the median value of $ \tau_{cr} $, that we denote as $\tilde{\tau_{cr}}$, since it is more representative in this case than the mean values. We conclude that history functions that end up in the limit cycle, on average, leave the chaotic attractor before and for a wider range of values of $ \tau $.

This type of distribution is in contrast with other analysis performed for non time-delayed systems. In \cite{Maslennikov2013,Maslennikov2018,Cantisan2020a} it was found that the values of the parameter for the transition follow a normal distribution instead of a gamma distribution. This might be one of the effects that delay causes in the system. It implies that there are some trajectories that tip at much larger values than the average.

\begin{figure}[h]
    \begin{center}
        \includegraphics[width=0.6\textwidth]{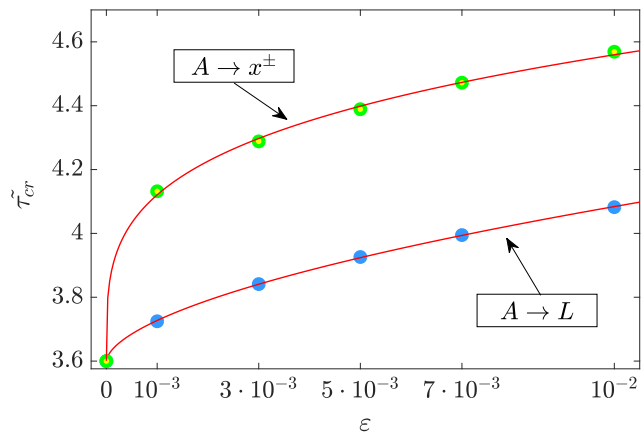}
    \end{center}
    \caption{Scaling law for the median value $\tilde{\tau_{cr}}$ with the parameter change rate. The points (in blue for trajectories that approach the limit cycle and in yellow/green the trajectories that approach the fixed points) correspond to numerically calculated values and in red we show a power law fit.}
    \label{scaling_law}
\end{figure}

If we repeat the same process for different values of $ \varepsilon $, we may deduce how it affects the value of $ \tau  $ for the transition. Figure \ref{scaling_law} depicts the dependence of $\tilde{\tau_{cr}}$ with $ \varepsilon $ for the trajectories that end up in the fixed points (yellow/green) and the limit cycle (blue). As we can see, for higher rates the delay phenomenon is more severe in both cases. However, we can see that on average the transition  occurs before for the trajectories heading to $ L $ rather than $ x^{\pm} $.

The points in figure \ref{scaling_law} are fitted to a power law that constitutes the numerical scaling law for the median value $\tilde{\tau_{cr}}$

\begin{equation}
    A \rightarrow x^{\pm}: \qquad \tilde{\tau_{cr}}  = a \cdot \varepsilon^{4/15} +\tau_{0}
    \label{scaling_eq}
\end{equation}
\begin{equation}
 A \rightarrow L: \qquad \tilde{\tau_{cr}}  = b \cdot \varepsilon^{4/7} +\tau_{0}
\end{equation}
where $ a,b >0 $ are constants and $ \tau_{0}=3.6 $ in our case,
with a $R$-square: $ R^{2}=0.9993 $ and $ R^{2}=0.9999 $
respectively. We have added the point for the limit $ \varepsilon
\rightarrow 0 $ which corresponds to the frozen-in case for which
the chaotic attractor looses stability at $ \tau=3.6 $.

This law indicates that an increasing parameter change rate
increases the parameter value for the transition. However, the
increase in $ \tilde{\tau_{cr}} $ is reduced for higher values of $
\varepsilon $ as the slope of the curves reduces with the parameter
change rate.

\section{Transient chaos interpretation} \label{sec5}

In this section we consider again our system with parameter drift
and the previous results from a different perspective. Until now, we
have considered that the drift in $ \tau $ represented a small
perturbation to the associated frozen-in system and that the chaotic
behavior after $ \tau=3.6 $ was a metastable state that ended at $
\tau_{cr} $. Also, we calculated the value of $ \tilde{\tau_{cr}} $
for which the delayed bifurcation takes place.

Now, we study the system without any previous knowledge of the
behavior of the frozen-in system. From a experimentalist point of view, sometimes it is the regime shift which indicates that there is a parameter drift and not the other way around. However, if this regime shift occurs later than expected due to the delay effect, the parameter may have reached dangerous values when it is noticed by the experimentalist \cite{Stevens2002,Wu2015}. This is why, in this section, we are interested in measuring the time that the system takes before the tipping.

In this time framework, we deal with a
non-autonomous system that behaves chaotically for a finite time
before reaching one of the attractors (see figure \ref{tau_cr}). In
this sense, the system presents transient chaos and the scaling law
predicts the end of the transient state.

\begin{figure}[h]
    \begin{center}
        \includegraphics[width=0.6\textwidth]{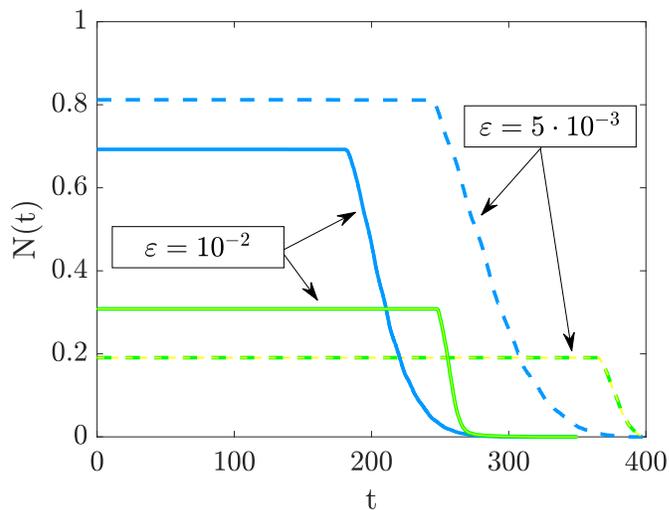}
    \end{center}
    \caption{Normalized number of trajectories that remain in the chaotic attractor for a time $t$. The dashed lines correspond to a parameter change rate of $ \varepsilon=5 \cdot 10^{-3} $ and the full ones to $ \varepsilon= 10^{-2} $. In blue the trajectories that approach the limit cycle and in yellow/green the ones that approach the fixed points. For slow parameter change rates the transient dynamics lasts for a longer period of time. This may create the false impression that the transient regime is the steady state.}
    \label{transient}
\end{figure}

In order to characterize the transient chaos regime, we
analyze the decay with time in the number of trajectories that still
present a chaotic behavior. In figure \ref{transient} we represent $
N(t) $ as the normalized number of trajectories in the chaotic
attractor for a time $ t $. The blue lines correspond to the
trajectories that end in the limit cycle and the yellow/green lines
to the ones that end in the $ x^{\pm} $ attractors. This is
calculated for $ \varepsilon=5 \cdot 10^{-3} $ (dotted lines) and $
\varepsilon=10^{-2} $  (full lines).

Note that the decay in $ \tau $ starts at $ t=100 $, and that $ \tau
$ and time $t$ are equivalent through equation (\ref{tau(t)}). When the curves
decrease to zero, the transient chaos regime ends and every
trajectory reaches its steady state.

As it can be seen, the decay with time slows down at the end of the
curve. This is related to the gamma distribution for $ \tau_{cr} $
(figure \ref{histogram}). In fact, we are representing nothing more
than the complementary cumulative distribution function of figure
\ref{histogram} in terms of time. There are some trajectories with a
transient lifetime higher than the average and this produces a delay
in reaching the steady state for the full set of history
functions. Similarly, the variance in figure \ref{histogram}
is reflected now as the time since the curve $ N(t) $ starts
decreasing until it reaches zero, this is more pronounced for
the set of trajectories going to the limit cycle (blue lines) while
the others present a more step-like shape.

If we compare between different values of $ \varepsilon $, we can assert by looking at figure \ref{transient} that the transient lifetime decreases with the change rate of the parameter shift. For faster changes in the parameter the steady state is reached faster, although it occurs for further values of $ \tau $ from $ \tau_{0} $.

\begin{figure}[h]
	\begin{center}
		\includegraphics[width=0.6\textwidth]{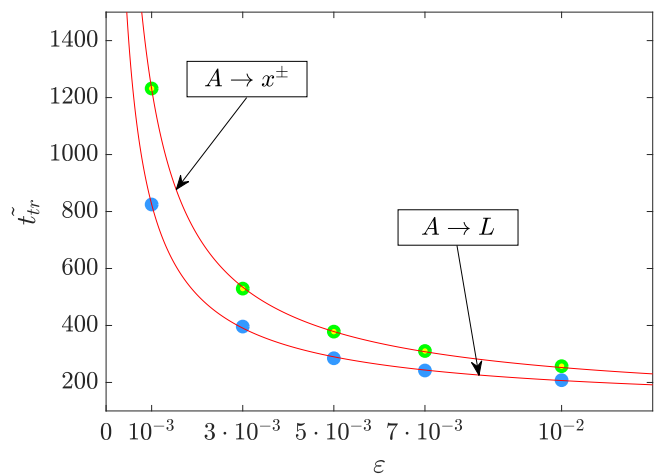}
	\end{center}
	\caption{Scaling law for the transient lifetime with the parameter change rate. The points (in blue for trajectories that approach the limit cycle and in yellow/green the trajectories that approach the fixed points) correspond to numerically calculated values and in red we show a power law fit.}
	\label{scaling_tr}
\end{figure}

Finally, we may translate the scaling law deduced in the previous section to a time framework just by using equation (\ref{tau(t)}):

\begin{equation}
        A \rightarrow x^{\pm}: \qquad \tilde{t_{tr}}  = c \cdot \varepsilon^{-13/15}+t_{1},
    \label{Scaling Law lifetime x}
\end{equation}

\begin{equation}
    A \rightarrow L: \qquad \tilde{t_{tr}}  = d \cdot \varepsilon^{-6/7}+t_{1},
    \label{Scaling Law lifetime L}
\end{equation}
where $\tilde{t_{tr}} $ refers to the median lifetime of the transient dynamics which follows the same gamma distribution as $ \tau_{cr} $ in figure \ref{histogram}. Also, $ t_{1} $ in our case is $ 100 $ and $ c, d $ are  positive constants. The numerically calculated points and the scaling laws are depicted in figure \ref{scaling_tr} for a visual guidance. 

Again, we obtain that the median value for the transient lifetime decreases with the parameter change rate following a power law. This decrease is more pronounced in the case of the trajectories that approach the fixed points as the slope is bigger in that case. For any change rate, the transient lifetime of the trajectories heading towards the limit cycle is shorter. Of course, when $ \varepsilon \rightarrow 0 $, the transient lifetime tends to infinity as in the limit when the parameter does not vary with time, the trajectories stay in the chaotic attractor forever. This is an interesting result as if a parameter is varying very slowly, a experimentalist may not notice that it is varying as the regime would not shift for a long time after the start of the drift. Specially, for the history functions corresponding to the tail of the lifetimes distribution. This may be dangerous in some engineering systems due to the parameter drift failure mechanism \cite{Wu2015,Stevens2002}.

\section{Conclusions} \label{sec6}
In the present work, we have analyzed the dynamics of a time-delayed
oscillator whose time delay suffers a drift in time. The time delay
increases linearly with time crossing a bifurcation value, after
which the system presents multistability. We have found that
trajectories initially in the chaotic attractor tip to one of the
remaining attractors with a certain probability that depends on the
parameter change rate. For faster rates more trajectories tip to the
fixed points $ x^{\pm} $ instead of the limit cycle. However,
predictability is lost in the non-autonomous system as the basins
are riddle-like. In this sense the chaotic attractor acts as a
memory-loss agent.

Also, we have found that the delay effect in the regime shift is
present in time-delayed systems. However, it acts in a different way
from the previously reported in other type of systems. The
distribution of values of $ \tau $ for which trajectories tip, this
is $ \tau_{cr} $, follows a gamma distribution instead of a normal
distribution. This implies that there is a set of trajectories that
tip at larger values of $ \tau $.

Furthermore, we derived two scaling laws relating the median value
of $ \tau_{cr} $, that is, $ \tilde{\tau_{cr}} $  and the parameter change rate for the cases for which the trajectories tip to the limit cycle and for the ones that
tip to the fixed points $ x^{\pm} $. We have also shown that
trajectories that end up in the limit cycle tip before than the ones
that tip to $ x^{\pm} $ and that for faster parameter change rates,
$ \tilde{\tau_{cr}} $ increases in both cases.

Finally, we have characterized the system behavior in terms of time.
As the chaotic dynamics lasts for a finite amount of time, we may
say that the system presents transient chaos and we derived a
scaling law for the transient lifetime. We conclude that for very small parameter drifts, the transient may last for unexpected long times. This may cause problems in the context of engineering due to  parameter drift failure mechanisms. On the other hand, for fixed parameter change rates, trajectories that end up in
the limit cycle have a shorter transient regime on average.

\FloatBarrier \ack This work has been supported by the Spanish State
Research Agency (AEI) and the European Regional Development Fund
(ERDF, EU) under Projects No.~FIS2016-76883-P and
No.~PID2019-105554GB-I00.

\section*{References}


\end{document}